\documentclass[reprint,superscriptaddress,amsmath,amssymb,aps,prl]{revtex4-1}

\usepackage{graphicx}
\usepackage{dcolumn}
\usepackage{bm}
\usepackage{color}
\usepackage{bbold}
\usepackage{soul}
\usepackage{placeins}%
\usepackage{epsfig,epic}
\newcommand{\ud}{\text{d}}
\providecommand*{\ui}{\text{i}}
\DeclareMathOperator{\arcosh}{arccosh}
\providecommand*{\map}{U}
\providecommand*{\K}{\kappa}
\providecommand*{\HrsActAng}{\mathcal{H}_{r:s}}
\newcommand{\Hrs}{H_{r:s}}
\providecommand*{\Ho}{\mathcal{H}_{0}}
\providecommand*{\Irs}{I_{r:s}}
\providecommand*{\Mrs}{M_{r:s}}
\providecommand*{\Vrs}{V_{r:s}}

\providecommand*{\Irat}{I_{\text{rat}}}
\providecommand*{\I}{I_{m}}
\providecommand*{\absreg}{\mathcal{L}}
\providecommand*{\Ich}{I_{\absreg}}
\providecommand*{\AT}{\mathcal{A}_{\text{T}}}
\providecommand*{\qmap}{\widehat{\map}}

\makeatletter
\let\Hy@backout\@gobble
\makeatother

\begin{document}

\title{Complex-Path Prediction of Resonance-Assisted Tunneling in Mixed
Systems}

\author{Felix Fritzsch}
\affiliation{Technische Universit\"at Dresden, Institut f\"ur Theoretische
             Physik and Center for Dynamics, 01062 Dresden, Germany}
\affiliation{Max-Planck-Institut f\"ur Physik komplexer Systeme, N\"othnitzer
Stra\ss{}e 38, 01187 Dresden, Germany}

\author{Arnd B\"acker}
\affiliation{Technische Universit\"at Dresden, Institut f\"ur Theoretische
             Physik and Center for Dynamics, 01062 Dresden, Germany}
\affiliation{Max-Planck-Institut f\"ur Physik komplexer Systeme, N\"othnitzer
Stra\ss{}e 38, 01187 Dresden, Germany}

\author{Roland Ketzmerick}
\affiliation{Technische Universit\"at Dresden, Institut f\"ur Theoretische
             Physik and Center for Dynamics, 01062 Dresden, Germany}
\affiliation{Max-Planck-Institut f\"ur Physik komplexer Systeme, N\"othnitzer
Stra\ss{}e 38, 01187 Dresden, Germany}

\author{Normann Mertig}
\affiliation{Technische Universit\"at Dresden, Institut f\"ur Theoretische
             Physik and Center for Dynamics, 01062 Dresden, Germany}
\affiliation{Max-Planck-Institut f\"ur Physik komplexer Systeme, N\"othnitzer
Stra\ss{}e 38, 01187 Dresden, Germany}
\affiliation{Department of Physics, Tokyo Metropolitan University,
Minami-Osawa, Hachioji 192-0397, Japan}

\date{\today}

\begin{abstract}
We present a semiclassical prediction of regular-to-chaotic tunneling in systems
with a mixed phase space, including the effect of a nonlinear resonance chain.
We identify  complex paths for direct and resonance-assisted tunneling
in the phase space of an integrable approximation with one
nonlinear resonance chain.
We evaluate the resonance-assisted contribution analytically and give a
prediction based on just a few properties of the classical phase space.
For the standard map excellent agreement with numerically determined
tunneling rates is observed.
The results should similarly apply to ionization rates and
quality factors.
\end{abstract}

\pacs{PACS here}

\maketitle

\begin{figure}[b]
\includegraphics{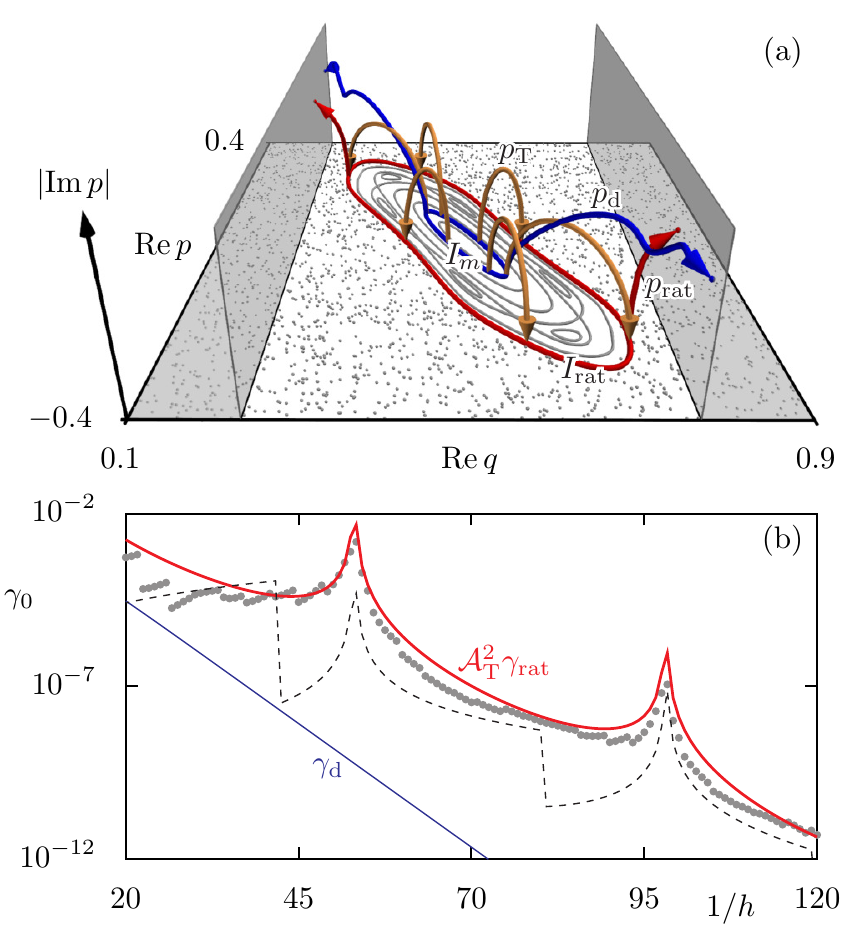}
\caption{\label{fig:gamma_ps}
(color online) (a) Phase space of the standard map for $\K=3.4$ with regular
tori (thin [gray] lines), and chaotic orbits ([gray] dots).
The shaded areas indicate a leaky region in the chaotic part
(see text).
The real tori and complex paths of the integrable approximation (thick, labeled
lines and arrows) connect the regular and the chaotic region.
(b) Tunneling rate $\gamma_0$ vs. inverse effective Planck
constant $1/h$.
Numerically obtained rates (dots) are
in excellent agreement with an analytic evaluation of the
resonance-assisted contribution $\AT^2\gamma_{\text{rat}}$.
Also shown are $\gamma_{\text{d}}$
and the perturbative result \cite{SchMouUll2011} (dashed line).
}
\end{figure}

Tunneling through energetic barriers is a textbook paradigm of quantum
mechanics.
While classically motion is confined to either side of the barrier, wave
functions exhibit contributions on both sides.
In contrast, nature often exhibits confinement on dynamically
disjoint regions of regular and chaotic motion in a mixed phase space, see
Fig.~\ref{fig:gamma_ps}(a).
Here, a classical particle follows a trajectory of regular motion while the
correponding wave function admits an exponentially small contribution on
the chaotic region.
This phenomenon is called dynamical tunneling \cite{DavHel1981, KesSch2011}.

Until today dynamical tunneling has emerged in many fields of physics.
It determines the vibrational spectrum of molecules \cite{DavHel1981},
ionization rates of atoms in laser fields \cite{WimSchEltBuc2006,ZakDelBuc1998},
and chaos-assisted tunneling oscillations \cite{LinBal1990, BohTomUll1993} in cold
atom systems \cite{Hen2001, SteOskRai2001}.
In optics dynamical tunneling is experimentally explored in microwave
resonators \cite{DemGraHeiHofRehRic2000, BaeKetLoeRobVidHoeKuhSto2008,
DieGuhGutMisRic2014, GehLoeShiBaeKetKuhSto2015} as well as
microlasers \cite{PodNar2005, ShiHarFukHenSasNar2010, ShiHarFukHenSunNar2011,
YanLeeMooLeeKimDaoLeeAn2010, KwaShiMooLeeYanAn2015, CaoWie2015, YiYuLeeKim2015,
YiYuKim2016}, where it determines the quality factor of lasing modes.
Here, a recent experimental breakthrough \cite{KwaShiMooLeeYanAn2015,
GehLoeShiBaeKetKuhSto2015} is the measured enhancement of dynamical tunneling
due to nonlinear resonance chains \cite{BroSchUll2001, BroSchUll2002}.

To reveal the universal features of dynamical tunneling it is extensively
studied theoretically \cite{ShuIke1995, ShuIke1998, PodNar2003, Kes2003,
PodNar2005, EltSch2005, Kes2005b, SheFisGuaReb2006, Kes2007, BaeKetLoeSch2008,
BaeKetLoeRobVidHoeKuhSto2008, ShuIke2008, ShuIshIke2008, ShuIshIke2009a,
ShuIshIke2009b, BaeKetLoeWieHen2009, BaeKetLoe2010, LoeBaeKetSch2010,
MerLoeBaeKetShu2013, HanShuIke2015, ShuIke2016, KulWie2016,
MerKulLoeBaeKet2016:p} mainly in model systems.
A central object is the tunneling rate $\gamma_m$, which describes the
transition from a state on the $m$th quantizing torus of the regular region into
the chaotic sea.
Qualitatively $\gamma_m$ can be understood from the theory of resonance-assisted
tunneling \cite{BroSchUll2001, BroSchUll2002, EltSch2005, SchMouUll2011}, see
dashed line in Fig.~\ref{fig:gamma_ps}(b):\
On average $\gamma_m$ decreases exponentially for decreasing wavelength
or decreasing effective Planck constant, i.e.~Plancks constant scaled to some typical action of the system.
In addition a drastic enhancement of $\gamma_m$ is observed for some values of
$h$.
This is due to resonant coupling of regular states, induced by a nonlinear
resonance chain \cite{Bir1913} within the regular region, see
Fig.~\ref{fig:gamma_ps}(a).

Despite extensive effort an intuitive, trajectory-based picture of dynamical
tunneling from regular to chaotic regions, including the effect of nonlinear
resonances is not yet available.
Semiclassical theories exist only for time-domain quantities \cite{ShuIke1995,
ShuIke1998}, cases when resonances are irrelevant \cite{MerLoeBaeKetShu2013},
and near-integrable systems \cite{Ozo1984, BroSchUll2002, DeuMouSch2013}.
On the other hand, quantitatively accurate predictions of $\gamma_m$
\cite{LoeLoeBaeKet2013, MerKulLoeBaeKet2016:p} explicitely require integrable
approximations \cite{BaeKetLoeSch2008, BaeKetLoe2010, LoeBaeKetSch2010,
KulLoeMerBaeKet2014} which needs some numerical effort.

In this paper we establish an intuitive, semiclassical, trajectory-based picture
of resonance-assisted regular-to-chaotic tunneling in systems with a mixed
phase space.
It results in a closed-form, analytic formula for tunneling rates $\gamma_{m}$.
Our approach gives excellent agreement with numerical results for the standard
map, which outperforms the perturbative approach see Fig.~\ref{fig:gamma_ps}(b).
Since our final formula requires just a few properties of the classical phase
space rather than the construction of a full integrable approximation, it should
also allow for estimating ionization rates and quality
factors and be helpful, e.\,g., for designing experimental setups.

\textit{Overview} ---
Our method is based on a semiclassical evaluation of a recently developed
non-perturbative prediction of $\gamma_{m}$ \cite{MerKulLoeBaeKet2016:p}.
At its heart is an integrable approximation of the regular region, which
includes the relevant nonlinear resonance chain \cite{KulLoeMerBaeKet2014}.
In that, we justify and generalize the use of semiclassical techniques developed
for near-integrable systems \cite{Ozo1984, BroSchUll2002, DeuMouSch2013} in the
wider class of generic systems with a mixed phase space.
In particular, the integrable approximation overcomes the separation of regular
and chaotic motion and allows for connecting real tori to the chaotic region via
tunneling paths through complexified phase space.
This gives the tunneling rate
\begin{align}
\gamma_m = \gamma_{\text{d}} + \AT^2 \gamma_{\text{rat}},
\label{eq:decay_chan}
\end{align}
which is composed of a direct contribution $\gamma_{\text{d}}$ and a
resonance-assisted contribution $\AT^2 \gamma_{\text{rat}}$, see
(blue and red) lines in Fig.~\ref{fig:gamma_ps}(b).
Figure~\ref{fig:gamma_ps}(a) gives an illustration of the
phase-space structures contributing to Eq.~\eqref{eq:decay_chan}:

(i) Quantizing torus $\I$ and direct tunneling paths $p_{\text{d}}$:
The quantizing torus $\I$, associated with the $m$th regular state, gives rise
to tunneling paths $p_{\text{d}}(q)$ with complex momentum emanating from the
turning points of $\I$.
See (blue) inner ring and arrows, respectively.
They connect $\I$ with the chaotic sea and determine the direct tunneling rate
$\gamma_{\text{d}}$, Eq.~\eqref{eq:gamma_drat_2}.

(ii) Partner torus $\Irat$ and resonance-assisted tunneling paths
$p_{\text{rat}}$:
A partner torus with action $\Irat$ on the opposite side of the nonlinear
resonance is connected with the chaotic sea by complex tunneling paths
$p_{\text{rat}}(q)$, see (red) outer ring and arrows.
They lead to the resonance-assisted tunneling rate $\gamma_{\text{rat}}$,
Eq.~\eqref{eq:gamma_drat_2}.

(iii) Tunneling paths $p_{\text{T}}$:
The tori $\I$ and $\Irat$ are connected by complex paths $p_{\text{T}}(q)$
bridging the resonance, see (orange) arrows.
They determine the tunneling amplitude $\AT$, Eq.~\eqref{eq:AT}.

\textit{Basic setting} ---
We derive our results
for kicked one-dimensional Hamiltonians
$H(q, p, t) = T(p) + V(q)\sum_n{\delta(t - n)}$.
For illustrations we use $T(p)=p^2/2$ and $V(q)=\K\cos{(2\pi q)}/(4\pi^2)$,
giving the paradigmatic standard map \cite{Chi1979},
which is widely used to study tunneling phenomena
\cite{EltSch2005, BaeKetLoeSch2008, LoeBaeKetSch2010, SchMouUll2011,
MerKulLoeBaeKet2016:p}.
At $\K=3.4$ the corresponding stroboscobic Poincar{\'e} map
exhibits a mixed phase space as shown in
Fig.~\ref{fig:gamma_ps}(a) with regions of regular motion (thin [gray] lines)
and
chaotic motion (dots).
It is governed by a regular island containing a prominent
$r$:$s$=6:2 nonlinear resonance chain and a surrounding chaotic sea.
Quantum mechanically the dynamics is given
by the unitary time-evolution operator
$\qmap = \exp{(-\ui V(\hat{q})/\hbar)}
\exp{(-\ui T(\hat{p})/\hbar))}$.
By introducing a leaky region $\absreg$ (shaded areas in
Fig.~\ref{fig:gamma_ps}(a)) close to the regular-chaotic border
we compute tunneling rates $\gamma_m$
as discussed in Ref.~\cite{MerKulLoeBaeKet2016:p}.
We focus on the ground state ($m=0$) which localizes on the innermost quantizing
torus of the regular island.
Its tunneling rate is shown in Fig.~\ref{fig:gamma_ps}(b) (dots).
Note that higher excited states ($m > 0$) show the same qualitative features.

\begin{figure}
\includegraphics{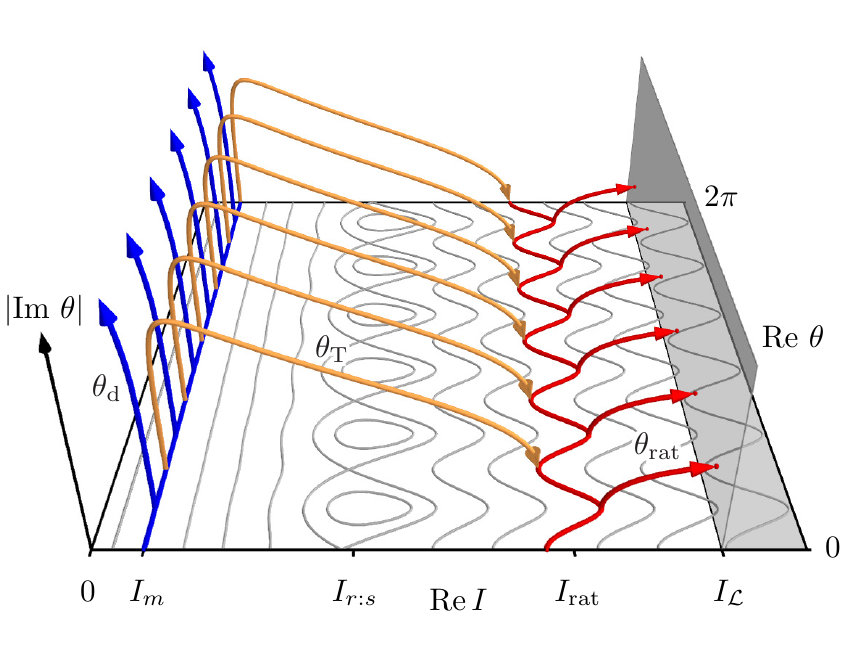}
\caption{(color online) Phase space of $\HrsActAng$ (thin [gray] lines) and
leaky region $\absreg$ (shaded area).
Real tori an complex paths (thick lines and arrows) are labeled in the figure.
}
\label{fig:ps_nf}
\end{figure}

\textit{Integrable approximation} ---
The key tool for deriving our prediction is an integrable approximation.
It is a one degree of freedom time-independent Hamiltonian,
which resembles the regular dynamics of the original system
\cite{KulLoeMerBaeKet2014}.
It is based on the universal description of the classical
dynamics in the vicinity of a $r$:$s$ resonance by the
pendulum Hamiltonian \cite{Ozo1984, BroSchUll2001, BroSchUll2002,
KulLoeMerBaeKet2014}
\begin{align}
\HrsActAng(\theta, I) = \Ho(I) + 2
\Vrs\left(\frac{I}{\Irs}\right)^{r/2}\cos{(r\theta)},
\label{eq:normalform}
\end{align}
using action-angle coordinates of $\Ho(I)$.
It is determined by the frequencies $\omega_0(I)$ of tori in the co-rotating
frame of the resonance, as $\omega_0(I) = \partial_I\Ho(I)$
\cite{KulLoeMerBaeKet2014}, where $\Ho(I)\approx(I - \Irs)^2/(2\Mrs)$ close to
the resonant torus $\Irs$.
The quantities $\Irs$, $\Vrs$, and $\Mrs$ can be computed from the position and
the size of the resonance chain and the linearized dynamics of its central
orbit \cite{EltSch2005, KulLoeMerBaeKet2014}.
The phase space is depicted by thin [gray] lines in Fig.~\ref{fig:ps_nf}.
Via a canonical transformation $T$ the Hamiltonian Eq.~\eqref{eq:normalform} is
mapped onto the phase space of the standard map giving $\Hrs(q, p) =
\HrsActAng(T^{-1}(q, p))$ \cite{KulLoeMerBaeKet2014}.
By quantizing and diagonalizing $\Hrs$ its eigenstates $\psi_m$ yield the
tunneling rate \cite{MerKulLoeBaeKet2016:p}
\begin{align}
\label{eq:FISA_semicl}
\gamma_m = \int_{\absreg}{\big | \psi_m(q) \big |^2 \ud q},
\end{align}
via the probability of $\psi_m(q)$ on the leaky region $\absreg$, which we
evaluate semiclassically in the following.
As discussed in Ref~\cite{MerKulLoeBaeKet2016:p},
Eq.~\eqref{eq:FISA_semicl} provides a good prediction of the tunneling rate $\gamma_m$ because $\psi_m(q)$ approximates the
corresponding state of the mixed system on the regular region and further
provides a sufficiently accurate extension into the regular--chaotic
border region, which dominates Eq.~\eqref{eq:FISA_semicl}.

\textit{WKB construction} ---
Using WKB-techniques \cite{BerMou1972, Cre1994} we now construct the state
$\psi_m$ within the integrable approximation $\Hrs(q,p)$.
This extends the semiclassical methods developed for integrable systems
\cite{DeuMouSch2013} to systems with a mixed phase space.
Note that, the use of the integrable approximation solves the problem of
natural boundaries \cite{GrePer1981, Per1982}.
Thus the integrable approximation is the key for connecting regular and
chaotic motion quantum mechanically.

Following \cite{Cre1994}, the wave function is constructed
from generalized plane waves with locally adapted momentum,
Eq.~\eqref{eq:psi_WKB}.
This requires the solutions $p_{\alpha}(q)$ of the equation $E_m = \Hrs(q,
p_{\alpha}(q))$, as depicted in Fig.~\ref{fig:gamma_ps}(a).
Here, $E_m$ is the energy of the wave function $\psi_{m}(q)$ obtained from EBK
quantization.
The position coordinate $q$ is real.
The real solutions $p_{\alpha}(q)$ describe the oscillatory
part of the wave function in classically allowed regions.
The complex solutions $p_{\alpha}(q)$ describe the
exponentially decreasing tunneling tails of the wave
function in classically forbidden regions.
In particular, they describe $\psi_{m}(q)$ in the leaky region, as required by
Eq.~\eqref{eq:FISA_semicl}.

Specifically, for resonances the geometry of paths gives the semiclassical wave
function as a superposition
\begin{align}
\psi_m(q) = \psi_{\text{d}}(q) + \AT\psi_{\text{rat}}(q).
\label{eq:psi_m}
\end{align}
Here, (i) $\psi_{\text{d}}(q)$ is the direct wave function, (ii)
$\psi_{\text{rat}}(q)$ is the resonant wave function, and (iii) $\AT$ is the
tunneling amplitude.
We now explain this in more detail:

(i) $\psi_{\text{d}}(q)$ describes the wave function along the quantizing torus
$\I$ in the classically allowed region of energy $E_m \approx
\Ho(I_m)$, which is obtained from EBK quantization of the torus $\I = \hbar(m + 1/2)$.
Using Airy-type connections \cite{Cre1994} this wave function is extended into
the classically forbidden region along the paths $(q, p_{\alpha}(q))$ (with
$\alpha=\text{d}$), see Fig.~\ref{fig:gamma_ps}(a), as
\begin{align}
  \label{eq:psi_WKB}
\psi_{\alpha}(q)  =\Big| \frac{\omega_0(\I)}
           {2\pi \partial_p \Hrs (q, p_{\alpha}(q))} \Big|^{1/2}
\!\!\!\exp{\left(\frac{\ui}{\hbar}\!\int^{q}{\!\!\!\!p_{\alpha}(\tilde{q})\ud
\tilde{q}}\right)}.
\end{align}
Here, $\omega_0(\I)/(2\pi)$ accounts for global normalization of the wave
function, while $1/|\partial_p \Hrs(q, p_{\alpha}(q))|$ is the classical probability
along $p_{\alpha}(q)$.
The complex action $\int{\!p_{\alpha}(q)\ud q}$, for which the lower limit
is one of the turning points on the torus $\I$, describes direct tunneling
into the leaky region.

(ii) Due to the presence of the nonlinear resonance there is an additional real
solution $\Irat$ with energy $E_m$ on the opposite side of the resonance chain.
Along this torus $\Irat$ we construct the wave function $\psi_{\text{rat}}(q)$.
In particular, the tunneling tails associated with the
solutions $p_{\text{rat}}(q)$ emanating from
$\Irat$ and connecting to the chaotic part of phase space, see
Fig.~\ref{fig:gamma_ps}(a), also obey Eq.~\eqref{eq:psi_WKB} with
$\alpha=\text{rat}$.
Note that $\omega_0(\I)/(2\pi)$ in Eq.~\eqref{eq:psi_WKB} must be kept for
normalization.
The lower limit of the action integral is one of the turning points on $\Irat$.

(iii) Finally, the tunneling amplitude is given by \cite{BroSchUll2002}
\begin{align}
\label{eq:AT}
\AT = \left|2 \sin{\left(\frac{\pi}{r\hbar}
                                   \left[\Irat - \I \right]\right)}\right|^{-1}
                \exp{\left(-\frac{\sigma}{\hbar} \right)},
\end{align}
where $\sigma = \text{Im} \int{\!p_{\text{T}}(q)\ud q}$ is the imaginary part of
the action of any path $p_{\text{T}}(q)$ connecting $\I$ to $\Irat$.
In particular, since there is no solution $p_{\alpha}(q)$ connecting $\I$ and
$\Irat$ along real positions, these paths are only sketched schematically
in Fig.~\ref{fig:gamma_ps}(a).
Note that this evaluation of $\AT$ based on paths with complex position is
formally beyond the WKB-construction used here.
It has been introduced and successfully applied for near-integrable systems in
Ref.~\cite{DeuMouSch2013}.
Further note that complex solutions which do not connect to a real torus are
neglected.

To summarize our construction, the wave functions $\psi_{\text{d}}(q)$ and
$\psi_{\text{rat}}(q)$, Eq.~\eqref{eq:psi_WKB}, together with the tunneling
amplitude, Eq.~\eqref{eq:AT}, give the wave function $\psi_{m}(q)$,
Eq.~\eqref{eq:psi_m}.
Inserting $\psi_{m}(q)$ into Eq.~\eqref{eq:FISA_semicl} and neglecting
interference allows for evaluating the integral in Eq.~\eqref{eq:FISA_semicl}
independently for $\psi_{\text{d}}(q)$ and $\psi_{\text{rat}}(q)$.
For solving these integrals we linearize the action integral in
Eq.~\eqref{eq:psi_WKB} around the boundary of the leaky region at $q_{\absreg}$.
We further account for the symmetry of the standard map with respect to the
central fixed  point.
This gives (i) the direct ($\alpha = \text{d}$) and (ii) the resonance-assisted
($\alpha = \text{rat}$) tunneling rate as
\begin{align}
  \label{eq:gamma_drat_2}
\gamma_{\alpha} &
= \frac{\hbar}{\text{Im}\,
p_{\alpha}(q_{\absreg})}|\psi_{\alpha}(q_{\absreg})|^2,
\end{align}
i.\,e., each tunneling rate is given by the value of the normalized WKB wave
function at the boundary $q_{\absreg}$ of the leaky region.

This construction constitutes our first main result.
Numerical evaluation of the semiclassically obtained tunneling rates shows
excellent agreement with numerical obtained tunneling rates (not shown).
This generalizes previous work, Refs.~\cite{Ozo1984, BroSchUll2002, DeuMouSch2013},
to the much larger class of mixed system, based on the powerful tool of integrable approximations.
It further provides a basis for a fully analytic prediction, which no
longer requires constructing integrable approximations explicitely:
Namely, we observe that for the standard map at $\K=3.4$ (and other
examples) the resonance-assisted contribution dominates the semiclassically
predicted decay rates for all values of the effective Planck constant.
In general, one can expect $\gamma_m = \AT^2\gamma_{\text{rat}}$, whenever the
resonance is sufficiently large, i.e. roughly speaking when it is visible within the
regular region.
The converse, that $\gamma_m$ is dominated by the direct tunneling rate
$\gamma_{\text{d}}$, may occur for small values of $1/h$ or if the resonance is
extremely small.

\textit{Analytic result }$\gamma_{m}\approx\AT^2\gamma_{\text{rat}}$ ---
In the following we derive an analytic formula which evaluates the dominating
term $\gamma_{m}=\AT^2\gamma_{\text{rat}}$ based on just a few properties of
the classical phase space.
To this end we use the pendulum Hamiltonian $\HrsActAng(\theta, I)$,
Eq.~\eqref{eq:normalform}, in action-angle coordinates of $\Ho(I)$ and the
action representation $\psi_m(I)$ of the WKB wave function, respectively.
This extends the WKB construction presented in Ref.~\cite{BroSchUll2002}
to the Hamiltonian~\eqref{eq:normalform}.
In this context the main novelty is to account both for the action dependence of the resonance term proportional to $\Vrs$ and to obtain a closed form
expression for $\gamma_{\text{rat}}$.

As a first approximation we extend the leaky region to all chaotic trajectories
as $I > \Ich$ (shaded area in Fig.~\ref{fig:ps_nf}).
In order to account for sticky motion, we choose $\Ich$ such that $2\pi\Ich$
is the area enclosing the regular region enlarged up to the most relevant
partial barrier \cite{EltSch2005, SchMouUll2011}.
While the basic features of $\gamma_{m}$ are preserved upon changing the leaky
region, it is worth noting that its details might change roughly up to two
orders of magnitude \cite{MerKulLoeBaeKet2016:p}.
This constitutes the main error of our prediction.

To construct the WKB wave function
the classical phase-space structures
$\theta_{\alpha}(I)$
fulfilling $\HrsActAng(\theta, I)=E_m$
for real actions
are required.
They are depicted in Fig.~\ref{fig:ps_nf} and obey
$\cos{(r\theta)} = \varphi(I)$, where
\begin{align}
\label{eq:phi}
\varphi(I) = \left(\frac{\Irs}{I}\right)^{r/2} \frac{E_m - \Ho(I)}{2\Vrs}.
\end{align}
Real solutions correspond to the tori oscillating around $\I$ and $\Irat$, i.e.
the classically allowed regions ([blue and red] thick lines) on opposite sides
of the resonance chain at $\Irs$.
We have $\I = \hbar(m + 1/2)$ and a reasonable approximation of $\Irat$ is
obtained from $\Irat \approx 2\Irs - \I + \left[E_m - \Ho(2\Irs -
\I)\right]/\omega_0(2\Irs - \I)$.
The torus $\I$ is accompanied by complex paths $\theta_{\text{d}}(I)$ ([blue]
arrows) which emanate from turning points with $I < \I$ and diverge at $I = 0$.
Furthermore, there are tunneling paths $\theta_{\text{T}}(I)$ ([orange] arrows)
with imaginary part $\text{Im}\, \theta_{\text{T}}(I) =
\arcosh{(|\varphi(I)|)}/r$
attached to turning points with $I > \I$ bridging the resonance towards $\Irat$.
Finally, there are complex paths $\theta_{\text{rat}}(I)$ ([red] arrows)
emanating from turning points with $I > \Irat$ on the partner torus.
They have imaginary part $\text{Im}\, \theta_{\text{rat}}(I) =
\arcosh{(|\varphi(I)|)}/r$ as well and connect $\Irat$ with the leaky region.

Using Eq.~\eqref{eq:psi_WKB}, with accordingly interchanged phase-space
coordinates, local WKB wave functions can be constructed from these paths.
Again a global construction of $\psi_m(I)$ is obtained by using Airy-type
connections at classical turning points \cite{Cre1994, BroSchUll2002}.
In action-angle coordinates the torus $\I$ is not directly connected with the
leaky region.
Thus, for $\psi_{m}(I)$ there is neither a direct contribution to the
WKB wave function within $\absreg$ nor a direct tunneling
rate $\gamma_{\text{d}}$ involved in this construction.
Consequently, one has $\psi_m(I) = \AT \psi_{\text{rat}}(I)$
within $\absreg$.
As the tunneling amplitude $\AT$, Eq.~\eqref{eq:AT}, is canonically invariant,
it can be computed in action-angle coordinates as well, requiring
the evaluation of $\sigma =
\text{Im} \int \theta_{\text{T}}\ud I$
from $\I$ to $\Irat$.
By approximating $\text{Im}\, \theta_{\text{T}}(I) \approx
\ln{(2|\varphi(I)|)}/r$, which is justified if $\Vrs \ll E_m$, and using only
the quadratic part of $\Ho(I)$, we find
\begin{align}
\label{eq:sigma}
\sigma = & \frac{\Irat - \I}{r}
 \ln\!{\left(\frac{(\Irat - \I)^2}
 {2e^{2}\Mrs \Vrs}\right)}  \nonumber \\
 & + \frac{\I}{2}\ln{\left( \frac{\I}{e\Irs}\right)}
-  \frac{\Irat}{2}\ln{\left( \frac{\Irat}{e\Irs}\right)},
\end{align}
which inserted in Eq.~\eqref{eq:AT} constitutes the first
part of our analytic expression.
Note that the first term coincides with the results obtained
for a simpler pendulum model in Ref.~\cite{BroSchUll2002}
while the remaining terms are related to the action
dependence of the resonance term proportional
to $\Vrs$ in Eq.~\eqref{eq:normalform}.

We proceed by computing $\gamma_{\text{rat}}$ by Eq.~\eqref{eq:FISA_semicl}
from the WKB wave function $\psi_{\text{rat}}(I)$ and
its probability inside the leaky region.
The WKB wave function $\psi_{\text{rat}}(I)$ is associated with
$\theta_{\text{rat}}(I)$ and computed analogously to Eq.~\eqref{eq:psi_WKB}
using action-angle coordinates.
Linearizing the tunneling action occurring in the exponential in
Eq.~\eqref{eq:psi_WKB} around $\Ich$ then gives
\begin{align}
\label{eq:gamma_rat_I2}
\gamma_{\text{rat}} & = \frac{r\hbar}{2\ln{(2|\varphi(\Ich)|)}}\big |
\psi_{\text{rat}}(\Ich) \big |^2,
\end{align}
which is in close analogy with Eq.~\eqref{eq:gamma_drat_2}.
Again the resonance-assisted tunneling rate is determined by the
normalized WKB wave function
\begin{align}
\label{eq:psi_rat_I2}
\big |\psi_{\text{rat}}(\Ich) \big |^2 =
\Big | \frac{\omega_0(\I)}{2r\pi\left(E_m - \Ho(\Ich) \right)}\Big |
\exp\!{\left(-\frac{2}{\hbar} \mathcal{S}_{\text{rat}} \right)}
\end{align}
at the boundary of the leaky region.
The tunneling action $\mathcal{S}_{\text{rat}}
= \text{Im} \int \theta_{\text{rat}}\ud I$ from $\Irat$ to $\Ich$
is evaluated similarly as Eq.~\eqref{eq:sigma} leading to
\begin{align}
\mathcal{S}_{\text{rat}} = & \frac{\Ich - \Irat}{r}
\ln\!{\left(\frac{(\Ich - \Irat)(\Ich - \I)}
                {2e^2\Mrs\Vrs}\right)}               \nonumber \\
&          + \frac{\Irat}{2}\ln\!{\left( \frac{\Irat}{e\Irs}\right)}
 - \frac{\Ich}{2}\ln\!{\left( \frac{\Ich}{e\Irs}\right)} \nonumber\\
&  + \frac{\Irat - \I}{r}\ln\!{\left(\frac{\Ich - \I}{\Irat - \I} \right)},
\label{eq:S_rat}
\end{align}
which concludes the computation of $\gamma_{\text{rat}}$.
\FloatBarrier

\textit{Discussion} ---
The evaluation of $\gamma_m\approx\AT^2\gamma_{\text{rat}}$, based on the
analytic expressions, Eqs.~\eqref{eq:AT} and \eqref{eq:sigma} for $\AT$ and
Eqs.~\eqref{eq:psi_rat_I2}--\eqref{eq:S_rat} for $\gamma_{\text{rat}}$, requires
just a few classical quantities, namely $\Irs$, $\Vrs$, and $\Mrs$ as well as
the frequencies $\omega_0(I)$.
This analytic prediction is in excellent agreement with numerically obtained
rates, see Fig.~\ref{fig:gamma_ps}(b).
The resonance peaks originate from the divergence of the prefactor $\AT$,
Eq.~\eqref{eq:AT}, i.\,e., they appear whenever $\Irat$ fulfills a quantization
condition $\Irat = \hbar(m + lr + 1/2)$.
In particular, at the resonance peak a hybridization between states associated
with the $m$th and $(m + lr)$th quantizing torus occurs.
This is the same resonance condition as obtained from perturbation theory
\cite{BroSchUll2001}.
In contrast, away from the resonance peak the tori $\I$ and $\Irat$ are still
energetically degenerate.
However, since $\Irat$ does not fulfill a quantization condition there is no
associated quantum state.
This is different from the perturbative framework, where several quantizing tori
of different energy contribute to the final prediction.
Finally, in contrast to perturbation theory, our result is dominated by a
single term for all values of the effective Planck constant.
Its overall exponential decay is dominated by the first term of the action
$\sigma$, Eq.~\eqref{eq:sigma}.
Hence, the slope of the exponential decay, as depicted in
Fig~\ref{fig:gamma_ps}(b), is roughly proportional to the width of the dynamical
tunneling barrier $\Irat - \I$.

\textit{Summary and outlook} ---
We derive a trajectory-based, semiclassical prediction of resonance-assisted
regular-to-chaotic tunneling rates in systems with a mixed phase space.
To this end we generalize the semiclassical picture valid in near-integrable
systems to the larger class of systems with a mixed phase space, based on
integrable approximations which include the relevant resonance chain.
From this result we find a direct and a resonce-assisted contribution.
The latter usually dominates the whole experimentally accessible regime of
large tunneling rates.
For this resonance-assisted contribution we derive a closed-form analytic
expression which depends on just a few properties of the classical phase space.
In particular, this expression does not require the explicit construction of
integrable approximations.
Testing our analytic result for the paradigmatic example of the standard map
we find excellent agreement with numerically determined tunneling rates.
We expect that our result should also apply to ionization rates and quality
factors.\\

We gratefully acknowledge fruitful discussions with
Yasutaka Hanada,
Kensuke Ikeda,
Julius Kullig,
Clemens L{\"o}bner,
Steffen L{\"o}ck,
Amaury Mouchet,
Peter Schlagheck,
and
Akira Shudo.
We acknowledge support by the Deutsche Forschungsgemeinschaft (DFG) Grant No.\
BA 1973/4-1.
N.M.\ acknowledges successive support by JSPS (Japan) Grant No.\ PE 14701 and
Deutsche Forschungsgemeinschaft (DFG) Grant No.\ ME 4587/1-1.
All 3D visualizations were created using \textsc{Mayavi}~\cite{RamVar2011}.

\end{document}